\documentstyle[times,graphics,epsfig,amsmath,amssymb,natbib]{mn2e}

\newcommand{\be}{\begin{equation}}
\newcommand{\e}{\end{equation}}
\newcommand{\bear}{\begin{eqnarray}}
\newcommand{\ear}{\end{eqnarray}}
\newcommand{\nline}{\nonumber \\}
\newcommand{\f}{\frac}
\newcommand{\de}{{\rm d}}
\newcommand{\del}{\partial}
\newcommand{\R}{{\cal R}}
\def\apj{ApJ}
\def\apl{ApJL}
\def\mnras{MNRAS}
\def\prd{PRD}

\begin{document}
  \title[Multi-frequency  angular power spectrum of 21 cm 
    signal]{The multi-frequency  angular power spectrum of 
the epoch of reionization   21 cm   signal}
\author[Datta, Choudhury \&  Bharadwaj ]
{Kanan K. Datta$^{1,2}$\thanks{E-mail: kanan@cts.iitkgp.ernet.in}, 
T. Roy Choudhury$^{1,3}$\thanks{E-mail: chou@ast.cam.ac.uk}
and
Somnath Bharadwaj$^{1,2}$\thanks{E-mail: somnathb@iitkgp.ac.in}\\
$^1$Centre for Theoretical Studies, IIT, Kharagpur 721302, India\\
$^2$Department of Physics \& Meteorology, IIT, Kharagpur 721302, India\\
$^3$Institute of Astronomy, Madingley Road, Cambridge CB3 0HA, UK}

\maketitle

\date{\today}

\begin{abstract}
Observations of  redshifted 21cm radiation from neutral hydrogen (HI) 
at high redshifts is an important future probe of reionization. We consider the  Multi-frequency Angular Power Spectrum 
(MAPS) to  quantify the statistics of the HI signal  as a joint 
function of the angular multipole $l$ and frequency separation $\Delta
\nu$. The signal at two different frequencies  is expected to
decorrelate as $\Delta \nu$ is
increased, and quantifying this 
is particularly important in deciding
the frequency resolution for future HI observations. This is also
expected to play a very crucial role in extracting 
the signal from foregrounds as the signal
is expected to decorrelate much faster than the foregrounds (which are
largely continuum sources) 
with increasing $\Delta \nu$. 

In this paper we develop formulae relating  MAPS to different
components of the three dimensional HI power spectrum taking into
account HI peculiar velocities. We show that the flat-sky
approximation provides a very good representation over the angular
scales of interest, and a final  expression  which is very simple to
calculate and interpret. 
We present results for
$z=10$ assuming a neutral hydrogen fraction of $0.6$
considering two models for the HI
distribution, namely, (i) DM: where the HI traces the dark matter and 
(ii) PR: where the effects of patchy reionization are incorporated
through two parameters which are  the bubble size and the clustering
of the bubble centers relative to the dark matter (bias) respectively.
We find that 
while the DM signal is largely featureless, the PR signal 
peaks at the angular scales of the individual bubbles where it is
Poisson fluctuation dominated, and  the  signal is considerably
enhanced for large bubble size. For most cases of interest at $l \sim
100$ the signal is uncorrelated beyond $\Delta \nu \sim 1 \, {\rm
  MHz}$ or even less, whereas this occurs around $\sim 0.1 \, {\rm
  MHz}$ at $l \sim 10^3$. The $\Delta \nu$ dependence also carries an
imprint of the bubble size and the bias, and is expected to be an
important probe of  the reionization scenario. Finally we find that the
$l$ range $10^3 - 10^4$ is optimum for separating out the cosmological
HI signal from the foregrounds, while this  will be extremely
demanding at $l<100$ where it is necessary to characterize the
$\Delta \nu$ dependence of the foreground MAPS to an accuracy  better
than $1 \%$.

\end{abstract}


\section{Introduction}
One of the major challenges in modern cosmology is to understand
the reionization history of the Universe. Numerous attempts 
have been made in this regard to constrain
the evolution of neutral hydrogen density (for recent reviews, see
\citealt{barkana,fck06,cf06}). Analyzes of
quasar absorption spectra suggest that almost all of the hydrogen
has been ionized around redshift $z < 6$
(\citealt{becker,fan}). A different set of constraints come from 
the Cosmic Microwave Background (CMB)
observations. The CMB photons scatter with free electrons (produced
because of reionization) which result in the suppression of 
the intrinsic anisotropies; at the same time a
polarization signal is generated from this scattering. 
Measurements of the electron scattering
optical depth
from recent CMB observations like WMAP (\citealt{spergel,page}) imply that
reionization started before $z \sim 10$. 
It thus seems from these two results
that reionization is an extended and complex process
occurring over a redshift range 6--15 (\citealt{cf06a,alvarez}). 
However, there exist limitations of using these observations 
to study the details of reionization process which would be occurring
at redshifts $\approx 10$. 
At present there is no indication of quasars at such high redshifts; on
the other hand, the electron scattering
optical depth is
sensitive only to the integral history of reionization and it may not be useful
to study the progress of reionization with redshift. 
In fact, it has been shown that the CMB polarization power
spectrum is weakly dependent on the details of the reionization
history (\citealt{kaplinghat,hu,haiman,colombo}), though
weak constraints could be obtained from  
upcoming experiments such as PLANCK\footnote{http://www.rssd.esa.int/Planck/}.

Perhaps the most promising 
prospect of studying reionization at various stages of its progress is 
through  future 21 cm observations(for recent review see \citet{furlanetto4}).   This basically involves
observing   the redshifted 21 cm line from the neutral hydrogen (HI)
at  high redshifts. The advantage of these experiments lie in the fact
that one can track neutral hydrogen fraction at any desired redshift
by appropriately tuning  the observation frequency. Tomography of the
HI distribution using observations of the redshifted 21 cm line is one
of the most  promising tool to study reionization.

The possibility of observing 21 cm emission from the cosmological
structure formation was first recognized by \citet{suny} and later studied
by \citet{hogan}, \citet{scott} and \citet{madau} considering both
emission and absorption against the CMB. 
More recently, the effect of heating of the
HI gas and its reionization on 21 cm signal has drawn great deal of attention
and has been studied in detail 
(\citealt{gnedin,shaver,tozzi,isfm,iliev,ciardi,furlanetto,miralda,chen,cooray1,cooray2,mcquinn,sethi,salvaterra,carilli}).

It is currently perceived that a statistical analysis of the
fluctuations  in the redshifted 21 cm signal which is present as a
minute component of the background radiation in all low frequency
radio  observations holds the greatest potential. 
This approach has been considered in the context of lower redshifts  
(\citealt{bharad1,bharad2,bharad3,bharad4}). A similar  formalism
can also be applied at high redshifts to probe reionization and also
the pre-reionization era  (\citealt{zal,furlanetto2,morales,
  bharad5,bharad6, bharad7,ali1,ali2,lz,he}).  One of the main aims of
epoch of reionization 21 cm studies is to probe the 
size, spatial distribution and evolution of the ionized  regions
which has been addressed by several groups 
(\citealt{furlanetto3,furlanetto1,wyithe,mellema,iliev1}). 

On the observational front, several initiatives
are currently underway.  The Giant 
Metre-Wave Radio Telescope
 (GMRT\footnote{http://www.gmrt.ncra.tifr.res.in};  
\citealt{swarup}) is already
functioning at several bands in frequency range 150-1420 MHz and can
potentially detect the 21 cm signal at high redshifts.
In addition, upcoming low-frequency experiments such as LOw Frequency
ARray (LOFAR\footnote{http://www.lofar.org/}), Square Kilometer Array
 (SKA\footnote{http://www.skatelescope.org/}), PrimevAl Structure
 Telescope (PAST\footnote{http://web.phys.cmu.edu/~past/}) have now
 raised the 
possibility to detect 21 cm signal from very high redshift and thus
motivated a detailed study of the expected 21 cm background  from
 neutral hydrogen at high redshifts. 

Although the redshifted 21 cm line can provide enormous amounts of
information, its detection is going to be a huge challenge. The 
signal is expected to be highly
contaminated by foreground radio emission. Potential 
sources for these foregrounds include synchrotron and free-free
emission from or Galaxy and external galaxies, low-frequency radio
point sources and free emission from electrons in the intergalactic
medium  (\citealt{shaver,dimat1,dimat,oh1,cooray3}).  
Since these foregrounds are much stronger than the 21 cm
signal arising from HI, it has been
suggested that it may be impossible to study reionization via
such observations (\citealt{oh}). 

However, there have been various proposals for tackling such
foregrounds, the most promising being the usage of
multi-frequency observations. 
It has been proposed that multi-frequency analysis of the 
radio signal can be useful in separating out the 
foreground (see e.g. \citealt{shaver,dimat1,gnedin2,dimat,santos}). 
In fact this has also been noted by most of the authors   who have 
studied the statistical signal from redshifted HI and have been
referred to earlier. 
The cross correlation between the HI
signal at different frequencies is expected to decay rapidly as the
frequency separation is increased (\citealt{bharad6,santos}) while 
the foregrounds 
are expected to have a continuum spectrum and hence should be correlated
across frequencies. This property of different foregrounds contaminants 
can in principle be used to remove them from expected 21 signal. 

In this paper, we develop the formalism to calculate the
multi-frequency angular power spectrum (hereafter MAPS) which can be
used to Analise the  21 cm   signal from HI  both in emission and 
absorption against the CMB. We restrict our attention
to HI emission which is the situation of interest for the epoch of
reionization.  In our formalism, we consider the  effect of
redshift-space distortions which has been  ignored in many of earlier
works. As noted by \cite{bharad5}, this is an important effect and can
enhance the mean signal by $50\%$ or more and the effect is expected
to be most pronounced in the multi-frequency analysis. 
We next use the flat sky
approximation to develop  a much simpler expression of MAPS which is
much easier to calculate and interpret than the   angular power
spectrum  written in terms of the spherical Bessel functions.   We
adopt a simple model for the HI distribution \citep{bharad6}
which incorporates patchy reionization  and use it to predict the
expected signal and study its multi-frequency properties. The model
allows us to vary properties like the size of the ionized regions and
their bias relative to the dark matter. We use MAPS to analyze the
imprint of these features on the HI signal and discuss their
implication for future HI observations.

As noted earlier, the HI signal at two different frequencies separated
by $\Delta \nu$ is expected to become uncorrelated as $\Delta \nu$ is
increases. As noted in \cite{bharad6},  the value of $\Delta
\nu$ beyond which  the signal ceases to be correlated depends on the
angular scales being observed and it is $< 1 \, {\rm MHz}$ in most
situations  of interest. A prior estimate of the multi-frequency
behavior is extremely important when planning HI observations. The
width of the individual frequency  channels sets the frequency
resolution over which the signal is averaged. This should be chosen 
sufficiently small so that the signal remains  correlated over  the
channel width. Choosing a frequency channel which is too wide would
end up averaging uncorrelated HI signal which would wash out various
important features in the signal, and also lead to a degradation in the
signal to noise ratio. In this context we also note that an earlier
work \citep{santos} assumed  individual frequency channels $1 \, {\rm
  MHz}$ wide and smoothed the signal with this before performing the
multi-frequency analysis. This, as we have already noted and shall
study in detail in this paper, is considerably larger than the $\Delta
\nu$ where the signal is uncorrelated and hence is  not the optimal
strategy for the analysis. We avoid such a pitfall by not
incorporating the finite frequency resolution of any realistic  HI
observations. It is assumed that the analysis of this paper be used to
determine the optimal  frequency channel width for future HI
observations. Further, it is quite straightforward to introduce a
finite frequency window into our result through a convolution.

The outline of this paper is as follows. In Section 2 we present the 
theoretical formalism for calculating MAPS of the expected 21 cm
signal considering the effect of HI peculiar velocity. The calculation
in the full-sky and the flat-sky approximation are both presented with
the details being given in separate Appendices. Section 2.1 defines
various components of the HI power spectrum and Section  2.4 presents
to models for the HI distribution. We use these models when making
predictions for the expected HI signal. We present our results in
Section 3 and also summarize our findings. In Section 4. we discuss
the implications  for extracting the signal from the
foregrounds. 
\section{Theoretical Formalism}

\subsection{The HI power spectrum}

The aim of this section is to set up the notation and calculate
the angular correlation $C_l$ for the 21cm brightness temperature
fluctuations. It is now well known (e.g.. \citealt{bharad6}) that the
excess brightness temperature 
observed at a frequency $\nu$ 
along a direction ${\bf \hat{n}}$ is given by
\be
T(\nu,{\bf \hat{n}}) 
= \bar{T}(z)~\eta_{\rm HI}\left(z, {\bf \hat{n}} r_{\nu}\right)
\label{eq:1}
\e
where the frequency of observation is related to the redshift by
$\nu = 1420/(1 + z)\, {\rm MHz}$. The quantity 
\be
r_{\nu} = \int_0^z \de z' \f{c}{H(z')}
\e
is the comoving distance
and 
\be
\bar{T}(z) \approx 25 {\rm mK} \sqrt{\frac{0.15}{\Omega_m h^2}}
~\left(\frac{\Omega_B h^2}{0.022}\right)
\left(\frac{1 - Y}{0.76}\right) \sqrt{\frac{1+z}{10}}
\e
where $Y \approx 0.24$ is the helium mass fraction and 
all other symbols have usual meaning. In the above relation, it
has been assumed that the Hubble parameter 
$H(z) \approx H_0 \Omega_m^{1/2} (1+z)^{3/2}$, which is a good
approximation for most cosmological models at $z > 3$.
The quantity $\eta_{\rm HI}$ is known as 
the ``21 cm radiation efficiency in redshift space'' (\citealt{bharad6}) and can be written 
in terms of the mean neutral hydrogen fraction $\bar{x}_{\rm HI}$ and the 
fluctuation in neutral hydrogen density field $\delta_{\rm HI}$ as
\bear
\eta_{\rm HI}(z,{\bf \hat{n}} r_{\nu}) &=& \bar{x}_{\rm HI}(z) 
[1 + \delta_{\rm HI}(z,{\bf \hat{n}} r_{\nu})]
\left(1 - \f{T_{\gamma}}{T_s}\right)
\nline
&\times&
\left[1 - \f{(1+z)}{H(z)} \f{\del v(z,{\bf \hat{n}} r_{\nu})}{\del
    r_{\nu}} \right] 
\ear
where $T_{\gamma}$ and $T_s$ are the temperature of the CMB and 
the spin temperature of the gas respectively. The term in the square 
bracket arises from the coherent components of the HI peculiar
velocities.

At this stage, it is useful to make a set of assumptions which
will simplify our analysis: (i) We assume that $T_s \gg T_{\gamma}$,
which corresponds to the scenario where the intergalactic gas is heated
well above the CMB temperature during the reionization process. 
This assumption is expected to 
be valid throughout the IGM soon after the formation of first
sources of radiation as  a substantial background of X-rays
from supernovae, star-formation, etc. can heat the IGM quickly. 
(ii) We assume that the HI peculiar velocity field is determined
by the dark matter fluctuations, which is reasonable as
the peculiar velocities mostly trace the dark matter potential wells.
We then have 
\be
\eta_{\rm HI}(z,{\bf \hat{n}} r_{\nu}) = 
\int \f{\de^3 k}{(2 \pi)^3} 
{\rm e}^{-{\rm i} k r_{\nu} ({\bf \hat{k} \cdot \hat{n}})} 
\tilde{\eta}_{\rm HI}\left(z, {\bf k}\right)
\label{eq:eta}
\e
where 
\be
\tilde{\eta}_{\rm HI}\left(z, {\bf k}\right) =\bar{x}_{\rm HI}(z)
 \left[\Delta_{\rm HI}(z,{\bf k})
+   ({\bf \hat{k} \cdot \hat{n}})^2
\Delta(z, {\bf k}) \right] \,,
\label{eq:eta_tilde}
\e
and $\Delta_{\rm HI}(z,{\bf k})$ and $\Delta(z,{\bf  k})$ are 
the Fourier transform of the fluctuations in the HI and the
dark matter densities respectively. Note  that $f(\Omega_m)$, which
relates peculiar velocities to the dark matter, has been assumed to
have a  value  $f(\Omega_m)=1$ which is reasonable  at the high  $z$
of interest here. 

For future use, we define the 
relevant three dimensional (3D) power spectra 
\bear
\langle \Delta(z, {\bf k}) \Delta^*(z, {\bf k'}) \rangle
&=& (2 \pi)^3 \delta_D({\bf k - k'}) P(z,k) 
\nline
\langle \Delta_{\rm HI}(z, {\bf k}) \Delta^*_{\rm HI}(z, {\bf k'}) \rangle
&=& (2 \pi)^3 \delta_D({\bf k - k'}) P_{\Delta^2_{\rm HI}}(z,k) 
\nline
\langle \Delta(z, {\bf k}) \Delta^*_{\rm HI}(z, {\bf k'}) \rangle
&=& (2 \pi)^3 \delta_D({\bf k - k'}) P_{ \Delta_{\rm HI}}(z,k) 
\ear
where $P(z,k)$ and $P_{\Delta^2_{\rm HI}}(z,k) $ are the power spectra
of the fluctuations in the dark matter and the HI densities
respectively, while  $P_{ \Delta_{\rm HI}}(z,k) $ is the
cross-correlation between the two.

\subsection{The multi-frequency angular power spectrum (MAPS)} 
The multi-frequency angular power spectrum of 21 cm  
brightness temperature fluctuations at two different frequencies
$\nu_1$ and $\nu_2$ is defined as
\be
C_l(\nu_1,\nu_2) \equiv
\langle a_{lm}(\nu_1) ~ a^*_{lm}(\nu_2) \rangle \,.
\e
In our entire analysis  $\nu_1$ and $\nu_2$ are assumed to differ by
only a small amount $\Delta \nu \ll \nu_1$, and it is convenient to
introduce the notation 
\be
C_l(\Delta \nu) \equiv C_l(\nu, \nu + \Delta \nu)
\e
where we do not explicitly show the frequency $\nu$ whose value will
be clear from the context. Further, wherever possible, we shall
not explicitly show the $z$ dependence  of various quantities like 
$\bar{T}$, $\bar{x}_{\rm HI}$, $P(k)$ etc., and it is
to be understood that these are to be evaluated at the appropriate 
redshift determined by $\nu$.

The spherical harmonic moment of $T(\nu,{\bf \hat{n}})$ are 
defined as
\bear
a_{lm}(\nu) &=& \int \de \Omega ~ Y^*_{lm}({\bf \hat{n}}) ~ 
T(\nu,{\bf \hat{n}}) 
\nline
&=&
\bar{T}
\int \de \Omega ~ Y^*_{lm}({\bf \hat{n}}) ~ 
\int \f{\de^3 k}{(2 \pi)^3} \tilde{\eta}_{\rm HI}\left({\bf k}\right)
{\rm e}^{-{\rm i} k r_{\nu} ({\bf \hat{k} \cdot \hat{n}})} \,.
\nline 
\label{eq:a_lm}
\ear
Putting the expression (\ref{eq:eta_tilde})
for $\tilde{\eta}_{\rm HI}\left({\bf k}\right)$ 
in the above equation, one can explicitly calculate the 
MAPS
 in terms of the three dimensional power spectra defined
earlier. We give the details of the calculation 
in Appendix \ref{sec:cl} and present only the final expression  for
the angular power spectrum at a frequency $\nu$ 
\bear
C_l(\Delta\nu)\!\!\!\!\!&=&\!\!\!\!\!
\f{2\bar{T}^2 ~ \bar{x}^2_{\rm HI}}{\pi}
\int_0^{\infty} k^2 \de k \,
\left[j_l(k r_{\nu}) j_l(k r_{\nu_2})
P_{\Delta^2_{\rm HI}}(k)
\right.
\nline
&-&\!\!\!\!\!
 \{j_l(k r_{\nu}) j''_l(k r_{\nu_2}) 
+ j_l(k r_{\nu_2}) j''_l(k r_{\nu})\}
P_{\Delta_{\rm HI}}(k)
\nline
&+&\!\!\!\!\! 
\left. j''_l(k r_{\nu}) j''_l(k r_{\nu_2}) P(k)
\right]
\label{eq:cl_delta_nu}
\nline
\ear
Here 
$j''_l(x)=\frac{d^2}{dx^2} j_l(x)$ 
and 
we have used the notation 
$r_{\nu_2} = r_{\nu} + r'_{\nu}~\Delta \nu$ 
with
\be
r'_{\nu} \equiv \f{\del r_{\nu}}{\del \nu} = -\f{c}{\nu_0} \f{(1+z)^2}{H(z)}.
\e
Note that equation (\ref{eq:cl_delta_nu}) predicts $C_l(\Delta \nu)$ from
the cosmological 21 cm  HI signal to be real. 

With increasing $\Delta \nu$, we expect the two spherical Bessel
functions $j_l(k r_{\nu_1})$ and  $j_l(k r_{\nu_2})$ 
to oscillate out  of phase. As a consequence the value of
$C_l(\Delta \nu)$ is expected to fall increasing $\Delta \nu$. 
We quantify this through  a dimensionless frequency
decorrelation  function defined as the ratio 
\be
\kappa_l(\Delta \nu) \equiv \f{C_l(\Delta \nu)}{C_l(0)} \,.
\label{eq:kappa_l}
\e
For a fixed multipole $l$, this fall in this function with increasing
$\Delta \nu$  essentially measures how quickly features at the  angular
scale $\theta \sim \pi/l$ in the 21 cm HI maps  at two different
frequencies become uncorrelated.  Note that  $0 \le
|\kappa_l(\Delta \nu)| \le 1$.

\subsection{Flat-sky approximation}
\label{sec:flat-sky}
Radio interferometers have a finite field of view which is determined  by
the parameters of the individual elements in the array. For example, 
at $150 \, {\rm MHz}$  this is around $3^{\circ}$ for the GMRT.
In most cases of interest it suffices to consider only   small
angular scales which correspond to $l \gg 1$. 
For the currently favored set of flat 
$\Lambda$CDM models, a 
comoving length scale  $R$ at redshift $z > 5$ would 
roughly correspond to a multipole
\be
l \approx 3 \times 10^4 \left(1 - \f{1.1}{\sqrt{1+z}}\right)
\left(\f{R}{h^{-1}\mbox{Mpc}}\right)^{-1}
\e
Thus, for length scales of $R < 100 h^{-1}$ Mpc
at $z \approx 10$, one would
be interested in multipoles $l > 200$.
For such high values of $l$ one can
work in the flat-sky approximation. 

A small portion of the sky can  be well
approximated by a plane. The unit vector 
${\bf \hat{n}}$ towards the direction of observation
can be decomposed as
\be
{\bf \hat{n}} = {\bf m} + \boldsymbol{\theta}; ~~
{\bf m} \cdot \boldsymbol{\theta} = 0 ; ~~ \mid \boldsymbol{\theta}
\mid \ll 1
\e
where ${\bf m}$ is a vector towards the center of the 
field of view and $\boldsymbol{\theta}$ is a two-dimensional
vector in the plane of the sky.
It is then natural to define
the two-dimensional Fourier transform of $T(\nu,{\bf \hat{n}})$
in the flat-sky as
\bear
\tilde{T}(\nu, {\bf U}) &\equiv& \int \de \boldsymbol{\theta} ~  
{\rm e}^{-2 \pi {\rm i} {\bf U \cdot} \boldsymbol{\theta}} ~
T(\nu,{\bf \hat{n}})  
\label{eq:T_four}
\ear
where $\bf U$, which corresponds to an  inverse angular scale, is the
Fourier space counterpart of $\boldsymbol{\theta}$. Using
equations  (\ref{eq:1}) and (\ref{eq:eta}),  and the fact that for the
flat-sky   we can approximate ${\bf k \cdot \hat{n}
  \approx  k \cdot \hat{m}}\equiv k_{\parallel}$ we have 
\be
\tilde{T}(\nu, {\bf U}) =\f{\bar{T}}{2 \pi \, r^2_{\nu}}  \int \de   
k_{\parallel} \, e^{i k_{\parallel}  r_{\nu}} \, \tilde{\eta}_{\rm
  HI}(k_{\parallel} {\bf \hat{m}}+ 2 \pi {\bf U}/r_{\nu}) \,.  
\e
It is useful to introduce  the $\tilde{\eta}_{\rm HI}$
power spectrum  $P_{\rm HI}$ defined as 
\be
\langle \tilde{\eta}_{\rm HI}({\bf k}) \tilde{\eta}_{\rm HI}({\bf
  k'}) \rangle = (2 \pi)^3 \delta^3_D({\bf k - k'}) P_{\rm HI}({\bf
  k}) \,.
\e
This is related to the other three power spectra introduced earlier
through 
\be
 P_{\rm HI}({\bf k})=\bar{x}^2_{\rm HI}(z) [ P_{\Delta^2_{\rm
       HI}}(k) 
+ 2  \mu^2  P_{\rm \Delta_{\rm HI}}(k) + \mu^4 
 P(k)]
\label{eq:PHI}
\e
where $\mu={\bf  \hat{m} \cdot \hat{k}}=k_{\parallel}/k$.
Note that the anisotropy of  $ P_{\rm HI}({\bf k})$ {\it i.e.,} 
its $\mu$-dependence arises  from  the peculiar velocities.

The quantities calculated  in the flat-sky approximation 
can be expressed in terms of their all-sky counterparts.
The correspondence between the all-sky angular power spectra and its 
flat-sky approximation is given by
\be
\langle \tilde{T}(\nu_1, {\bf U}) \tilde{T}^*(\nu_2, {\bf U'}) \rangle
= 
C_{2 \pi U}(\nu_1, \nu_2) ~ \delta^{(2)}_D({\bf U - U'})
\label{eq:T_Cl}
\e
where $\delta^{(2)}_D({\bf U - U'})$ is the two-dimensional 
Dirac-delta function. The details of the above calculation
are presented in Appendix \ref{sec:flat-all}.
Thus allows us to estimate the angular power spectrum
$C_l$ under the flat-sky approximation which has 
a much simpler expression (\citealt{bharad6})
\be
C^{\rm flat}_l(\Delta \nu)
=
\f{\bar{T}^2~ }{\pi r_{\nu}^2}
\int_{0}^{\infty} \de k_{\parallel} \, 
\cos (k_{\parallel} r'_{\nu} \Delta \nu) \, P_{\rm HI}({\bf k})
\label{eq:fsa} 
\e
where the vector ${\bf k}$ has magnitude  $k = \sqrt{k_{\parallel}^2 +
  l^2/r^2_{\nu}}$ {\it ie.} ${\bf k}$ has  components $k_{\parallel}$
and $l/r_{\nu}$  along the line of sight and  in the plane of the sky
respectively. 
It is clear that the angular power spectrum $C_l(\Delta \nu)$
is calculated by summing over all Fourier modes ${\bf k}$ whose
projection in the plane of the sky has a magnitude $l/r_{\nu}$. We
also see that $C_{l}$ is  determined by the power spectra only for
modes $k \ge l/r_{\nu}$. 

The flat-sky angular power spectrum $C^{\rm flat}_l(0)$ is
essentially the 2D power spectrum of the HI distribution on a plane 
at the distance $r_{\nu}$ from the observer, and for $\Delta \nu=0$
equation (\ref{eq:fsa}) is just the relation between the 2D power spectrum
and its 3D counterpart (\citealt{peacock}). For $\Delta \nu \neq 0$ it is
the cross-correlation of the 2D Fourier components of the HI
distribution on two different planes, one at $r_{\nu}$ and another at
$r_{\nu+\Delta \nu}$. Any 2D Fourier mode is calculated from its full  3D
counterparts by projecting the 3D modes onto the  plane where the 2D
Fourier mode is being evaluated. The same set of 3D modes contribute
with different phases when they are projected onto two different
planes. This gives rise to  $\cos(k_{\parallel} r'_{\nu} \Delta \nu)
$ in equation (\ref{eq:fsa}) when the same 2D mode on two different planes
are cross-correlated and this in turn causes the  decorrelation of 
$C^{\rm   flat}_l(\Delta \nu)$  with increasing $\Delta \nu$. 

Testing the range of $l$ over which the flat-sky approximation is
valid, we find that for the typical HI power spectra 
$C^{\rm flat}_l(\Delta \nu)$ is in agreement with  the full-sky
$C_l(\Delta \nu)$  calculated using  equation (\ref{eq:cl_delta_nu})
at a level better than 1 per cent  for  angular modes $l > 10$.
Since the integral in equation (\ref{eq:fsa}) is much simpler
to compute, and   more straightforward to interpret, we use 
the flat-sky approximation of $C_l$ for our calculations in the rest
of this paper.

Note that equation (\ref{eq:fsa}) is very
similar to the expression for the visibility correlations 
[equation (16) of \citealt{bharad6}] expected in radio
interferometric observations of redshifted 21 cm HI emission. 
The two relations differ only in a  proportionality factor which 
incorporates the parameters of the telescope being used for the
observation. This reflects the close relation between the visibility
correlations, which are the directly measurable quantities in radio
interferometry, and the $C_l$s considered  here.

\subsection{Modeling the  HI distribution}

The crucial quantities in calculating the angular correlation function
are the three dimensional power spectra $P(k)$,
$P_{\Delta_{\rm HI}}(k)$ and 
$P_{\Delta^2_{\rm HI}}(k)$.
The form of the 
dark matter power spectrum $P(k)$ is relatively
well-established, particularly within the linear theory. We shall be
using the standard expression given by \citealt{bunn}.

The power spectrum of HI density fluctuations $P_{\Delta^2_{\rm
    HI}}(k)$ and its cross-correlation with the dark matter 
fluctuations $P_{\Delta_{\rm HI}}(k)$  are both largely unknown,
and determining these is one of the most important aims of the future  
redshifted 21 cm observations. We adopt  two
simple models with a few parameters which capture the  salient
features of the HI distribution.

The first model, which we shall denote as DM,  assumes homogeneous
reionization where the HI 
traces the dark matter , i.e., $\Delta_{\rm HI} = \Delta$.
In this model we have 
\be 
P_{\Delta^2_{\rm HI}}(k)=P_{\Delta_{\rm HI}}(k) =
P(k)
\e
which we use in equation (\ref{eq:cl_delta_nu}) to
calculate $C_l(\Delta \nu)$. Alternately, we have 
\be 
P_{\rm HI}(k)=\bar{x}^2_{\rm HI} \,
(1 + \mu^2)^2 \,P(k)
\label{eq:P_HI_DM}
\e
which we can use in equation (\ref{eq:fsa}) to
calculate $C_l(\Delta \nu)$ in the flat-sky approximation. 
This model has only one free parameter namely the mean neutral
fraction $\bar{x}_{\rm HI}$.

The second model, denoted as PR, 
incorporates  patchy reionization. It is assumed that
reionization occurs through the growth of completely ionized regions 
(bubbles) in the hydrogen distribution. The bubbles are assumed to be
spheres,  all with the same   
comoving radius $R$,  their centers tracing the dark matter
distribution with a possible bias $b_c$.  While in reality there will
be a spread  in the  shapes and sizes of the ionized patches, we
can consider $R$ as being the  characteristic size at any particular
epoch. The  distribution of the centers of the ionized regions
basically incorporates the fact that the ionizing sources are expected
to reside  at the peaks of the dark matter density distribution and
these are expected to be strongly clustered. 
For non-overlapping  spheres  the fraction of ionized volume is given
by 
\be
\bar{x}_{\rm HII} \equiv 1 - \bar{x}_{\rm HI}
= \f{4 \pi R^3}{3} \tilde{n}_{\rm HI}
\e
where $\tilde{n}_{\rm HI}$ is the mean comoving number density of
ionized spheres and we use $\R$ to denote the ratio 
$\R=\bar{x}_{\rm HII}/\bar{x}_{\rm HI}$. 
This model has been discussed in  detail 
in Bharadwaj and Ali (2005), and we have 
\be
\Delta_{\rm HI}({\bf k}) =
\left[1 -   b_c \R 
 W(k R)\right] 
\Delta({\bf k})
- \R
W(k R) \Delta_P({\bf k})\,.
\e
The HI fluctuation is a sum of  two parts, one which is correlated
with the dark matter distribution and an uncorrelated Poisson 
fluctuation $\Delta_P$. The latter  arises from the discrete nature of
the HII regions and has a  power spectrum $\tilde{n}_{\rm HI}^{-1}$. 
Also,  $W(y) = (3/y^3) [\sin y - y \cos y]$ is the 
spherical top hat window function arising from the Fourier transform
of the spherical  bubbles. This gives  
\be 
P_{\Delta^2_{\rm HI}}(k)=
\left[1 -  \R b_c 
 W(k R)\right]^2 P(k)+ 
\f{[\R W(k R)]^2}{\tilde{n}_{\rm    HI}}
\e
and 
\be
P_{\Delta_{\rm HI}}(k) = \left[1 -  \R b_c  W(k R)\right] P(k)
\e
which we use in equation (\ref{eq:cl_delta_nu})  to
calculate $C_l(\Delta \nu)$. Alternately, we have 
the HI power spectrum (Bharadwaj and Ali, 2005) 
\be
P_{\rm HI}({\bf k})=\bar{x}^2_{\rm HI} \left\{[1-\R b_c W(k R)  +
  \mu^2]^2 
P(k) + \f{[\R W(k R)]^2}{\tilde{n}_{\rm    HI}} \right\}
\label{eq:P_HI_patchy}
\e
which we can use in equation (\ref{eq:fsa}) to calculate $C_l(\Delta \nu)$
in the flat-sky approximation. 

 Our analysis assumes non-overlapping spheres and
hence it is valid only when a small fraction of the HI is ionized
 and the bias is not very large. As a consequence we restrict these
 parameters to the range $\bar{x}_{\rm   HI} \ge 0.5$ and
 $b_c \leq 1.5$. We note that in the early stages of reionization 
({\it ie.} $\bar{x}_{\rm HII} \ll  1$) equation (\ref{eq:P_HI_patchy})
matches the  HI power spectrum  
 calculated by \cite{wh05}, 
though their  method of arriving at the final result is
somewhat  different and  is quite a bit  more involved.

For large values of the bubble size ($R \ge 8$ Mpc) the 
power spectrum is essentially determined by the Poisson 
fluctuation term. At  length-scales larger than the bubble size  ($k <
\pi/R$) we have  $W^2(k R) \approx 1$, and hence the power spectrum 
is practically constant.   Around scales corresponding to the
characteristic 
bubble size $k \approx \pi/R$, the window function $W(kR)$ starts
decreasing which introduces a prominent drop in $P_{\rm HI}({\bf k})$.
For smaller length-scales ($k > \pi/R$),
the power spectrum shows oscillations arising from the  window
function $W(k R)$. At sufficiently large $k$ 
the power spectrum  $P_{\rm HI}({\bf k})$  is  dominated by the dark
matter fluctuations and it approaches the DM model. 
We should mention  here that the
oscillations in $P_{\rm HI}({\bf k})$ are a consequence of  the
ionized bubbles being  spheres, all of the same 
size.   In reality there will be  a spread in the
bubble shapes and sizes, and it is quite likely that such
oscillation will  be washed out (\citealt{wh05})
but we expect the other features of
the PR model discussed above to hold if the characteristic  bubble size
is large $(R \ge 8 \, {\rm Mpc})$. 

For smaller values of $R$, the power spectrum $P_{\rm HI}({\bf k})$ 
could be dominated either by the term containing the dark matter
power spectrum $P(k)$ or by the Poisson fluctuation
term, depending on the value of $\R b_c = 2 b_c/3$.

In addition to the effects considered above, 
the random motions within clusters
could significantly modify the signal by
 elongating  the HI clustering pattern along the line of sight   
[the Finger of God (FoG) effect].   We have incorporated this effect 
by multiplying the power spectrum $P_{\rm HI}({\bf k})$ with  an  extra 
Lorentzian term $(1+k^2_{\parallel} 
 \sigma_P^2/a^2 H^2)^{-1}$ (\citealt{sheth,ballinger}) where $\sigma_P$ is the one
 dimensional pair-velocity dispersion in relative galaxy velocities.

The cosmological parameters used
throughout this paper  are  those determined as the best-fit values by
WMAP 3-year data release,  i.e.,  
$\Omega_m=0.23, \Omega_b h^2 = 0.022, n_s = 0.96, h = 0.74, \sigma_8
= 0.76$ (\citealt{spergel}). Further, without any loss of generality, 
we have restricted our analysis to a single  redshift $z = 10$ which
corresponds to a frequency $\nu = 129$ MHz,  and have assumed 
$\bar{x}_{\rm HI} = 0.6$ (implying $\R = 2/3$) which is consistent
with currently favored reionization models.

The effects of patchy reionization are expected to be most prominent
at around $z \approx 10$ in currently favored reionization models.
At higher redshifts, the reionization is in its preliminary stages
($\bar{x}_{\rm HII} \ll 1$) and
the characteristic bubble size $R$ is quite  small. We expect the DM
model to hold at this  epoch.  At lower redshifts the HI signal
is expected to be drastically diminished because most of the
hydrogen would be ionized.
Given this, it is optimum to study the HI signal properties 
at some intermediate redshift where $\bar{x}_{\rm HII} \sim 0.5$
and $R$ is reasonably large. For the currently favored
reionization scenarios, it seems that these properties
are satisfied at $z \approx 10$ (\citealt{cf06a}), 
which we shall be studying in the
rest of this paper.

\section{Results}

\begin{figure}
\includegraphics[width=0.45\textwidth]{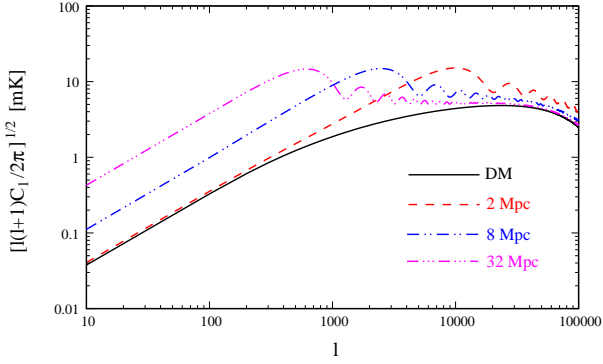}
\caption{The angular power spectrum of   HI brightness temperature
  fluctuations for different models of the HI distribution assuming
  $b_c=1$ and $\sigma_p = 0$. } 
\label{fig:cl}
\end{figure}

We first consider the angular power spectrum $C_l(\Delta \nu)$ at
$\Delta \nu=0$ for which  the results are shown in Figure  \ref{fig:cl}.
As discussed earlier,  $C_l(0)$ is essentially the 2D power spectrum
of HI 
fluctuations evaluated at the 2D Fourier mode $l/r_{\nu} \approx l \times
10^{-4} {\rm Mpc}^{-1}$.  
The results for the DM model serve as the fiducial case against which
we compare different possibilities for patchy reionization.

\begin{figure*}
\rotatebox{270}{\includegraphics[width=0.7\textwidth]{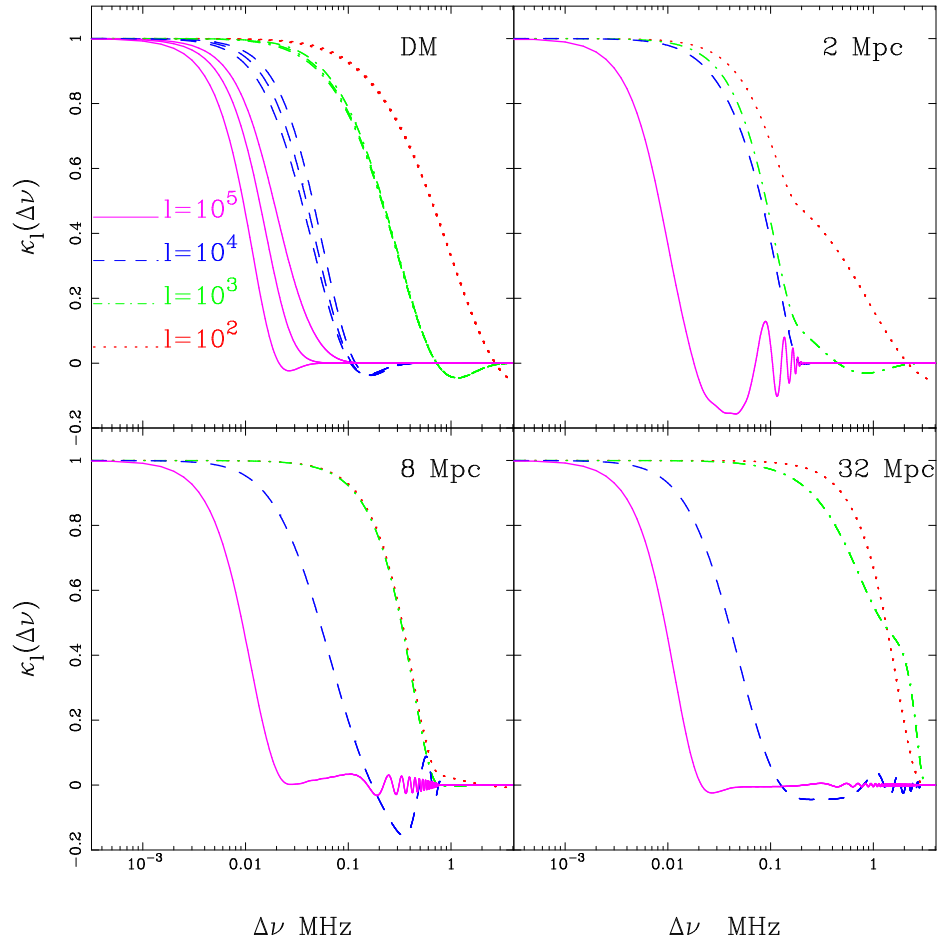}}
\caption{The frequency decorrelation function
$\kappa_l(\Delta \nu)$ [defined in equation (\ref{eq:kappa_l})] 
at $l=10^2,10^3,10^4,10^5$. Results are shown for the DM model and the
PR model with $b_c=1$ and the $R$ values shown in the figure. For the
DM model, we show results incorporating the 
FoG effect using $\sigma_P= 20 \, {\rm and } \, 40 \, {\rm km/s}$. 
For each $l$ value $\kappa_l(\Delta \nu)$ decreases  faster for
$\sigma_P=0$ and slowest for $\sigma_P=40\, {\rm km/s}$. There is a
significant  change due to the 
FoG effect only at $l \ge 10^{4}$. 
}
\label{fig:kappa_l}
\end{figure*}

 For large bubble size ($R \ge 8~\mbox{Mpc}$)  the HI 
signal is  dominated by  Poisson fluctuations and it is well described
through 
\begin{equation}
\sqrt{l (l+1) C_l} \propto \sqrt{x_{\rm HII}} \, \bar{T} \,
 \frac{R}{r_{\nu}} \,  l
\label{eq:scl}
\end{equation}
 on scales larger  than the bubble.  At these angular scales  
the HI signal is substantially enhanced compared to  the DM model. 
 For smaller bubble size, the large
angle signal is sensitive to the bias $b_c$. The signal is very
similar to the DM model for $b_c=0$ and it is suppressed for higher
bias. In all cases (large or small bubble size), the signal is Poisson
fluctuation dominated on scales comparable to the bubble size and it 
 peaks at $l \approx \pi r_{\nu}/R$, with no dependence on $b_c$. The
 HI signal traces the dark matter on scales which are much smaller
 than the bubble size.    

We next consider the behavior of $\kappa_l(\Delta \nu)$, the
frequency decorrelation function shown in  Figure \ref{fig:kappa_l}. 
For the DM model (upper left panel) where the HI fluctuations trace
the dark matter we 
find that the frequency difference $\Delta \nu$ over which the HI
signal remains correlated reduces monotonically  with increasing $l$.  
For example, while for $l=100$ $\kappa_l(\Delta \nu)$ falls to
$\sim 0.5$  at $\Delta \nu \sim 500 \, {\rm KHz}$, it
occurs much faster ($\Delta \nu \sim 10 \, {\rm KHz}$) for $l=10^5$. 
Beyond the first zero crossing $\kappa_l(\Delta \nu)$ becomes
negative (anti-correlation) and exhibits  a few highly damped
oscillations very close to zero. These oscillations arise from the
$\cos$ term in equation (\ref{eq:fsa}). 
The change in the
behavior of $\kappa_l(\Delta \nu)$ for the DM model arising from 
the FoG  effect is also shown in the same panel.
\citet{wh05} have proposed that $\sigma_p$ is
expected to have a value $\sim 30 \, {\rm km/s}$ at $z \sim 8$;  
in view of this, 
we show results for  $\sigma_p=20 \, {\rm and} \, 40 \,   {\rm km/s}$.  
We find that there is a discernible change at $l \ge 10^{4}$, and the
FoG effect causes the signal to remain correlated for a larger value
of $\Delta \nu$.  For $\sigma_p=20 \, {\rm km/s}$, 
the change is at most $15 \%$ for $l=10^{4}$ and
around $100 \%$ at $l=10^5$. Though we have not shown it explicitly, 
we expect similar changes due to FoG effect in the PR model also.

\begin{figure*}
\includegraphics[width=84mm]{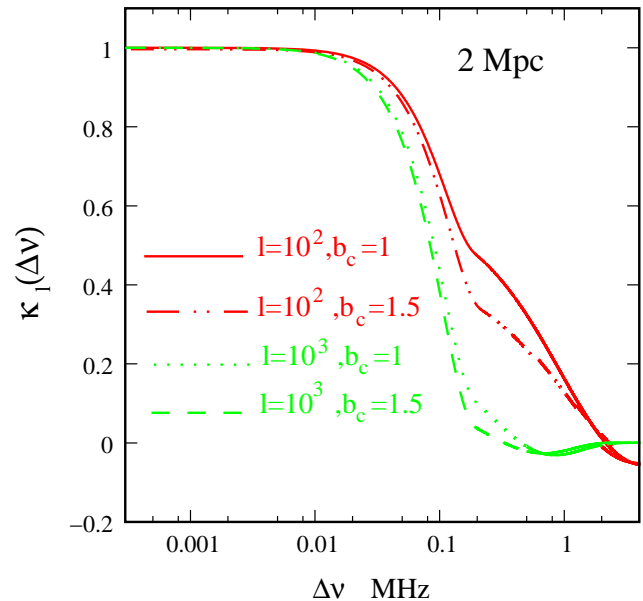}
\caption{This shows the $b_c$ dependence of the frequency decorrelation
  function $\kappa_l(\Delta \nu)$ for the PR model with $R=2 \, {\rm Mpc}$.}
\label{fig:kl2_bias}
\end{figure*}

The patchy reionization model shows distinct departures from the DM
model in the behavior of $\kappa_l(\Delta \nu)$. This reflects the
imprint of the bubble size and the bias on the $\Delta \nu$
dependence. For $R=2 \, {\rm Mpc}$ and $b_c=1$ (upper right panel) the
large $l$ ($l > 1000$, comparable 
to bubble size) behavior is 
dominated by the Poisson fluctuation of the individual bubbles which
makes $\kappa_l(\Delta \nu)$ quite distinct from the DM model. Notice
that for $l=10^3$, $\kappa_l(\Delta \nu)$ falls faster than the DM
model whereas for $l=10^4$ it falls slower than the DM model causing
the $l=10^3$ and $10^4$  curves to nearly overlap.  
 The oscillations seen in $C_l$ as a function 
of $l$ in Figure \ref{fig:cl} are also seen in the $\Delta \nu$
dependence of $\kappa_l(\Delta \nu)$ at large $l$ ($10^5$).  
The behavior at $l=10^2$ is a combination of the dark matter and the
ionized bubbles, and is sensitive to $b_c$. For $b_c=1$, the initial
decrease in  $\kappa_l(\Delta \nu)$ is much steeper  than the DM model
with a sudden break after which the curve flattens. Figure
\ref{fig:kl2_bias} shows the $b_c$ dependence for $l=10^2$ and $10^3$. 
The bias dependence is weak for $l=10^3$ where the Poisson
fluctuations begin to dominate. For $l=10^2$, changing $b_c$ has a
significant affect only near the break in $\kappa_l(\Delta \nu)$
leaving much of the curve unaffected. For a smaller bubble size we
expect a behavior similar to $R=2 {\rm Mpc}$, with the $b_c$
dependence being somewhat more pronounced and the Poisson dominated
regime starting from a larger value of $l$.

For large bubble size $(R \geq 8 {\rm Mpc})$ the large angle
HI signal ($l < \pi r_{\nu}/R$) is entirely determined by the Poisson
fluctuations where the  
signal is independent of $l$. This is most clearly seen for $R=8 {\rm
  Mpc}$ where the $\kappa_l(\Delta \nu)$ curves for
$l=10^2$ and $l=10^3$  are identical.  For both $R=8 {\rm Mpc}$ and
$32 \, {\rm Mpc}$ the large $l$ behavior of $\kappa_l(\Delta \nu)$
approaches that of the DM model. 

\begin{figure*}
\includegraphics[width=.45\textwidth]{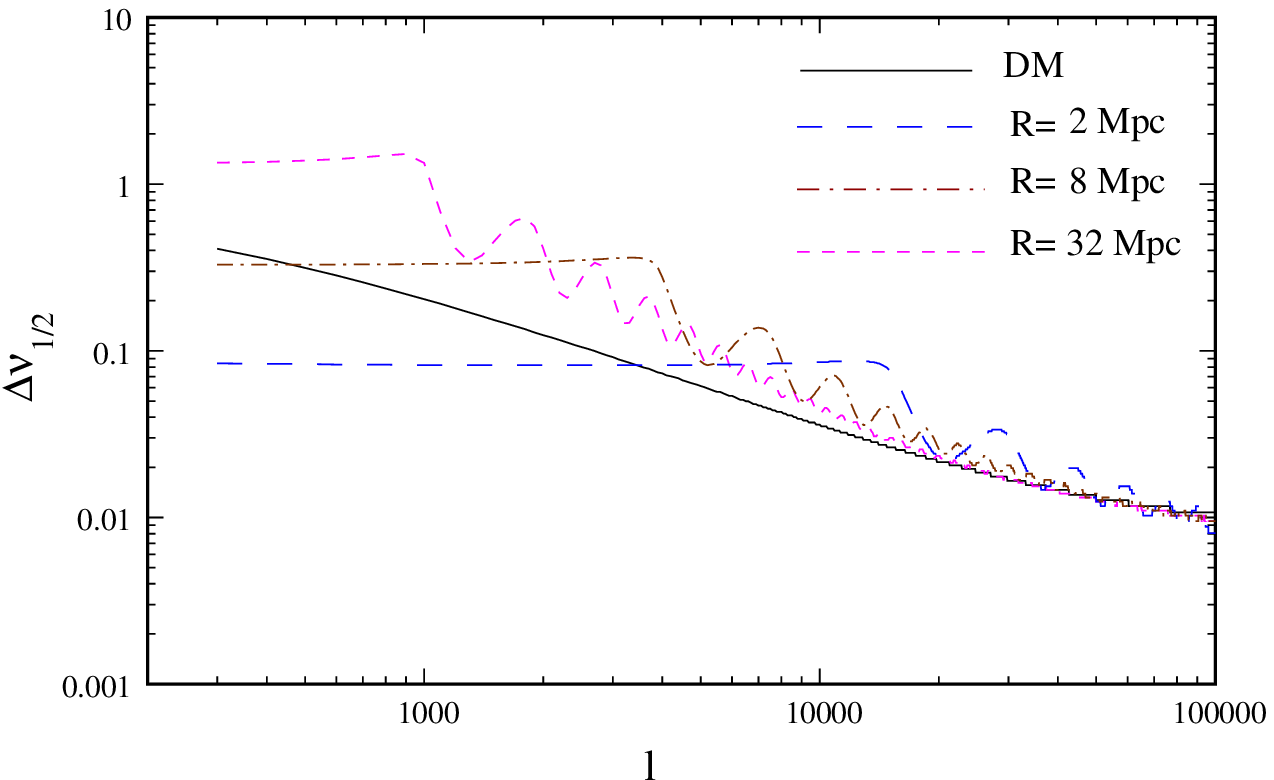}
\caption{This shows $\Delta \nu_{1/2}$ vs. $l$ for the DM model and
  the PR  model with $b_c=1$ for the $R$ values shown in the figure.  }
\label{fig:ldel}
\end{figure*}

In the final part we quantify the frequency difference $\Delta \nu$
across  which the HI signal at two different frequencies remain
correlated.   To be more precise, we study  the behavior  of   
$\Delta \nu_{1/2}$  which is defined such that $\kappa_l(\Delta
\nu_{1/2})=1/2$ {\it ie.} the correlation falls to  50\% of  
its peak value at $\Delta \nu=0$. We study this for different angular
scales (different $l$) for the various models of HI distribution
considered here. 
The main aim of this exercise is to determine the  frequency
resolution that would be required to   study the HI fluctuations on a
given angular scale $l$. Optimally one would like to use a frequency
resolution smaller than $\Delta \nu_{1/2}$. A wider frequency channel
would combine different uncorrelated signals whereby the signal would
cancel out. Further, combining such signals would not lead to an
improvement in the signal to noise ratio. Thus it would be fruitful to
combine the signal at two different frequencies only as long as they
are correlated and not beyond, and we use $\Delta \nu_{1/2}$ to
estimate this.   
The plot of $\Delta \nu_{1/2}$ vs $l$ for the different HI 
models is shown in Figure \ref{fig:ldel}.

We find that for the DM model $\Delta \nu_{1/2}$ falls monotonically
with $l$ and the relation  is  well approximated by a power law
\begin{equation}
\Delta \nu_{1/2}= \, 0.2 {\, \rm MHz} \times
\left( \frac{l}{10^3} \right)^{-0.7} 
\end{equation}
which essentially says that $\Delta \nu_{1/2}\sim 0.66 \, {\rm MHz}$ on
$1^{\circ}$ angular scales,  $\Delta \nu_{1/2}\sim 0.04 \, {\rm MHz}$
on $1'$ angular scales and $\Delta \nu_{1/2}\sim 2 \, {\rm KHz}$
on $1''$ angular scales.

For the PR model, as discussed earlier, $\kappa_l(\Delta \nu)$ is $l$
independent on  angular scales larger than the bubble size $l < \pi
r_{\nu}/R$. As a consequence $\Delta \nu_{1/2}$ also is independent of
$l$ and it depends only on the bubble size $R$. This can be well
approximated by   
\be
\Delta \nu_{1/2} \approx 0.04~\mbox{MHz}~ \left(\f{R}{\mbox{Mpc}}\right)
\e
which given a large value $1.3 \, {\rm MHz}$ for $R=32 \, {\rm Mpc}$
while it falls  below the DM model for  $0.08 \, {\rm MHz}$ for $R=2
\, {\rm   Mpc}$. The large $l$ behavior of $\Delta \nu_{1/2}$
approaches the DM model though there are oscillations which persist
even at large $l$. 

We note that our findings are consistent with the earlier findings of
\citet{bharad6} whereas they significantly different from the 
results  of \citet{santos} who assume frequency channels of  $1
\, {\rm MHz}$ which is too large.

\section{Implications for separating signal from Foregrounds}
\begin{figure*}
\includegraphics[width=.45\textwidth]{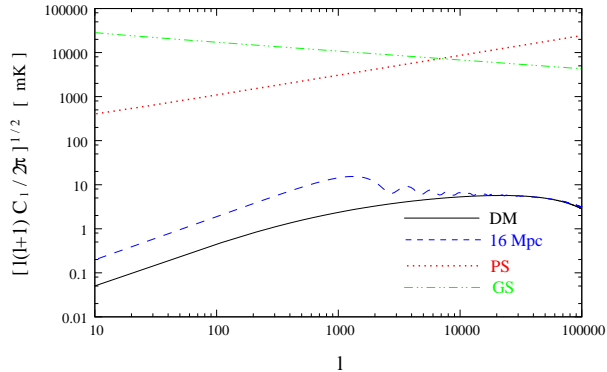}
\caption{Angular power spectrum $C_l(0)$ at $\nu=129 \,  {\rm  MHz}$
  for the two most dominant foreground components, the   diffuse
  galactic synchrotron radiation (GS) and the extragalactic point  
  sources (PS) assuming $S_{cut}=0.1 \, {\rm mJy}$. The expected
  signal is also shown  for the DM model and the PR model with $R= 16
  \,   {\rm Mpc}$. }

\label{fig:cl_fg}
\end{figure*}

Astrophysical foregrounds are expected to be several order of
magnitude stronger than the 21 cm signal. The MAPS foreground
contribution at a frequency $\nu$ can be parametrized as \citep{santos} 
\be
C_l(\Delta \nu)=A \left(\frac{\nu_f}{\nu} \right)^{ \bar{\alpha}}
\left(\frac{\nu_f}{\nu+\Delta \nu} \right)^{ \bar{\alpha}}
\left(\frac{1000}{l}\right)^{\beta}  I_l(\Delta \nu)
\label{eq:fg}
\e
where  $\nu_f=130$ MHz and $\bar{\alpha}$ is the mean spectral index. The actual spectral
index varies with line of sight across the sky and this causes the
foreground contribution to decorrelate with increasing frequency
separation $\Delta \nu$ which is quantified through the foreground
frequency 
decorrelation function $I_l(\Delta \nu)$  \citep{zal}
 which has been  modeled as 
\be
I_l(\Delta \nu)=\exp\left[ - \log_{10}^2 \left(1 +\frac{\Delta \nu}{\nu}
 \right)/2   \xi^2 \right] \,.
\e 
\begin{figure*}
\includegraphics[width=0.9\textwidth]{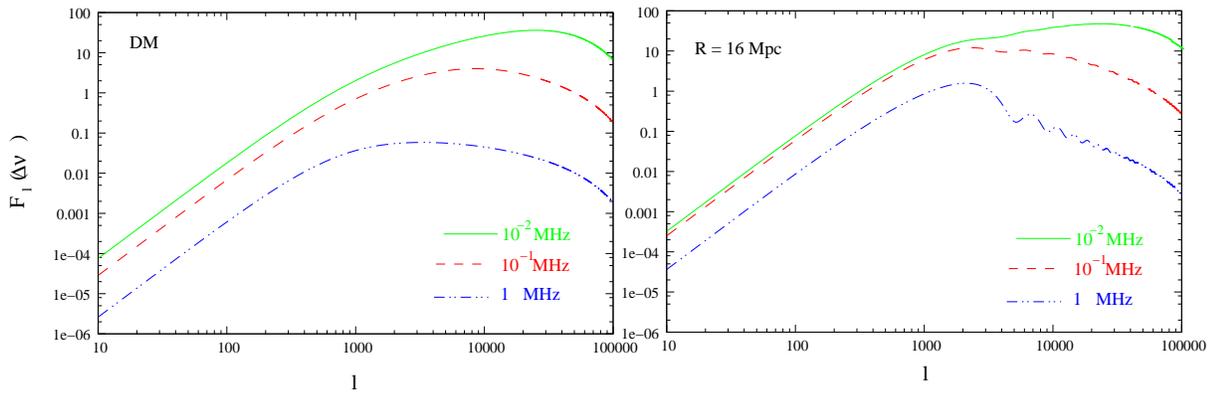} 
\caption{This shows $F_l(\Delta \nu)$ (defined in equation \ref{eq:Fl})
for the $\Delta \nu$ values shown in the figure. We consider both 
the DM model(left panel) and the PR model(right panel).}
\label{fig:Fl}
\end{figure*}

We consider the two most dominant  foreground
components namely extragalactic point sources and the 
diffuse synchrotron radiation from our own galaxy.  Point sources above
a flux level $S_{cut}$ can be identified in high-resolution images and
removed. We assume $S_{cut}=0.1 {\rm mJy}$ and adopt the parameter
values from  Table~1  of \citet{santos} for $A$, $\bar{\alpha}$,
$\beta$ and $\xi$. Figure \ref{fig:cl_fg} shows   the expected
$C_l(0)$ for the  signal and 
foregrounds. The galactic synchrotron radiation dominates at large
angular scales $l<10,000$ while the extragalactic point sources
dominate at small angular scales. For all values of $l$, the 
foregrounds are at least two orders of magnitude larger than the
signal.  

The foregrounds have a continuum spectra, and the contributions 
at a frequency separation  $\Delta \nu$  are expected to be
highly correlated. For $\Delta \nu=1 \, {\rm MHz}$, the foreground
decorrelation function $I_l(\Delta \nu)$ falls by only $2 \times
10^{-6}$ for the galactic synchrotron radiation and by $3 \times
10^{-5}$ for the point sources. In contrast, the HI decorrelation
function $\kappa_l(\Delta \nu)$ is nearly constant at very small
$\Delta \nu$ and then has a sharp drop well within  $1 \, {\rm MHz}$, and 
is largely uncorrelated  beyond. 
This holds the promise of allowing  the signal to be separated from
the foregrounds.  A possible  strategy 
is to cross-correlate different frequency channels of    the full data
which has both  signal and foregrounds, and to  
use the distinctly different  $\Delta \nu$ dependence to separate the
signal from the foregrounds  \citep{zal}. An alternate approach 
is to subtract  a best fit continuum spectra along each line of sight
\citep{wt06} and then determine the  power spectrum. 
 This is expected to be an  effective foreground subtraction method
 in data  with very  low noise levels. 
We consider the former 
approach here, and discuss the implications of our results.  

MAPS characterizes the joint $l$ and $\Delta \nu$ dependence 
which is expected to be different for the signal and the 
foregrounds. For a fixed $l$,  it will be possible to separate the
two with relative ease  at a frequency separation $\Delta \nu$ if  the
decrement in  the signal $C_l(0)[1-\kappa_l(\Delta \nu)]$  is more
than that of  the foregrounds $C_l(0)[1- I_l(\Delta \nu)]$.  
Note that because the foregrounds are much stronger  than the
HI signal, a very small decorrelation of the foreground
contribution  may cause a decrement in $C_l(\Delta \nu)$  which is
larger than 
that due to the  signal. 
We use  $F_l(\Delta \nu)$ defined as the ratio of the two
decrements 
\be 
F_l(\Delta \nu)=\frac{\{C_l(0) [1-\kappa_l(\Delta \nu)]\}_{\rm Signal}}
{\{C_l(0) [1-I_l(\Delta \nu)]\}_{\rm Foregrounds}}
\label{eq:Fl}
\e

to asses the feasibility of separating the HI signal from the
foregrounds. This gives an estimate of the
accuracy at which the $\Delta \nu$  dependence of the foreground 
 $C_l(\Delta \nu)$ has to be  characterized  for the
signal to be detected. 
 Note  we assume  that the  
$\left(\frac{\nu_f}{\nu+\Delta \nu} \right)^{ \bar{\alpha}}$ 
term in eq. (\ref{eq:fg}) can be  factored out before considering the
decrement in the  
foreground. Figure \ref{fig:Fl} shows the results for the DM model and
the PR model with $R=16 \, {\rm Mpc}$. 
First we note that $F_l(\Delta \nu)$ peaks at the angular scales
corresponding to $l \sim 10,000$  $(ie. \, 2^{'})$ and the prospects
of separating the signal from the foregrounds are most favorable  
at these scales. A detection will be possible in the range 
$l>1000$, $\Delta \nu \le 10 {\rm KHz}$ and $l>400$,$\Delta \nu
\le 100 {\rm KHz}$ for the DM and PR models respectively
provided  the  $\Delta \nu$ dependence of the
foregrounds $C_l(\Delta \nu)$ can be characterized with an
uncertainty less than order unity. The $l$ and $\Delta \nu$ range
would  increase if the $\Delta \nu$ dependence of the foreground
$C_l(\Delta \nu)$ were characterized to $10 \%$ accuracy. The largest
angular scales ( $l<100$ ) would require an accuracy better than  $1
\%$ which would possibly set the limit for forthcoming observations. 

The angular modes  $l=1,000$ and $l=10,000$ correspond to baselines
with antenna separations of  $\sim 300 \, {\rm  m}$  and 
$\sim 3 \, {\rm km}$ respectively. This baseline range is quite well
covered by the GMRT, and also the forthcoming interferometric
arrays.  This is possibly the optimal range for a detection. 
A possible detection  strategy would be to use  the $\Delta
\nu$ behavior of $C_l(\Delta \nu)$ in the range where $F_l(\Delta
\nu) \ll 1$ to characterize the foreground contribution. This can be
extrapolated to predict the foreground contribution at small $\Delta
\nu$ and any excess relative to this prediction can be interpreted as
the HI signal. A very precise determination of the
$\Delta \nu$ dependence of the foreground contribution would require a
very large $\Delta \nu$ range in the region where $F_l(\Delta \nu) \ll
1$, and a bandwidth of $\sim 10 {\rm MHz}$ would be appropriate. On the
other hand, at $l \sim 10,000$ the HI  $C_l(\Delta \nu)$ 
decorrelates  within $\sim 50 {\rm KHz}$ [or equivalently $F_l(\Delta
\nu)$ shows a considerable drop between $10 \, {\rm KHz} $ and $100 \,
   {\rm KHz} $, see Figure \ref{fig:Fl}], and it would be desirable to
   have   a frequency resolution better than  $\sim 10 {\rm KHz}$ 
 to optimally differentiate  between the signal and   the
 foregrounds. A lower resolution of $\sim 20 \, {\rm KHz}$ would
 possibly suffice at $l \sim 1,000$, particularly if the PR model
 holds.

\section*{Acknowledgment}
The authors would like to thank Sk. Saiyad Ali for useful discussions.
KKD is supported by a junior research fellowship of Council of 
Scientific and Industrial Research (CSIR), India.

\appendix

\section{Calculation of the angular power spectrum $C_{\lowercase{l}}(\nu_1,\nu_2)$}
\label{sec:cl}

In this section, we present the details of the calculation for
the 21 cm angular power spectrum $C_l(\nu_1,\nu_2)$. The first
step would be to calculate the spherical harmonic component $a_{lm}$
of the brightness temperature $T(\nu, {\bf \hat{n}})$. 
Using the expression (\ref{eq:eta_tilde}) 
for $\tilde{\eta}_{\rm HI}\left({\bf k}\right)$ in 
equation (\ref{eq:a_lm}), the expression for
$a_{lm}$ can be written as
\bear
a_{lm}(\nu) 
&=& 
\bar{T} ~ \bar{x}_{\rm HI}
\int \de \Omega ~ Y^*_{lm}({\bf \hat{n}}) ~ 
\int \f{\de^3 k}{(2 \pi)^3}
\nline
&\times&
\left[\Delta_{\rm HI}({\bf k}) 
+ ({\bf \hat{n} \cdot \hat{k}})^2 \Delta({\bf k})\right]
{\rm e}^{-{\rm i} k r_{\nu} ({\bf \hat{k} \cdot \hat{n}})}
\ear
which then essentially involves solving angular integrals of the forms
$\int \de \Omega ~ Y^*_{lm} ~ 
{\rm e}^{-{\rm i} k r_{\nu} ({\bf \hat{k} \cdot \hat{n}})}$
and $\int \de \Omega ~ Y^*_{lm} ~ ({\bf \hat{n} \cdot \hat{k}})^2
{\rm e}^{-{\rm i} k r_{\nu} ({\bf \hat{k} \cdot \hat{n}})}$ respectively.
Expanding the term 
${\rm e}^{-{\rm i} k r_{\nu} ({\bf \hat{k} \cdot \hat{n}})}$
in terms of spherical Bessel functions $j_l(kr_{\nu})$, one can show that
\be
\int \de \Omega ~ Y^*_{lm}({\bf \hat{n}}) ~ 
{\rm e}^{-{\rm i} k r_{\nu} ({\bf \hat{k} \cdot \hat{n}})}
=
4 \pi (-{\rm i})^l j_l(k r_{\nu}) Y^*_{lm}({\bf \hat{k}})
\e
Differentiating the above equation with respect to $k r_{\nu}$ twice
\be
\int \de \Omega ~ ({\bf \hat{k} \cdot \hat{n}})^2 ~ Y^*_{lm}({\bf \hat{n}}) ~ 
{\rm e}^{-{\rm i} k r_{\nu} ({\bf \hat{k} \cdot \hat{n}})}
= -4\pi
(-{\rm i})^l ~ j''_l(k r_{\nu}) Y^*_{lm}({\bf \hat{k}})
\e
where $j''_l(x)$ is the second derivative of $j_l(x)$ 
with respect to its
argument, and can be obtained through the recursion relation
\bear
(2 l + 1) j''_l(x) \!\!\!\!\! &=& \!\!\!\!\!\f{l(l-1)}{2 l -1} j_{l-2}(x)
-\left[\f{l^2}{2 l - 1} + \f{(l+1)^2}{2 l +3}\right] j_l(x)
\nline
\!\!\!\!\!&+&\!\!\!\!\!
\f{(l+1)(l+2)}{2 l +3} j_{l+2}(x)
\ear
So the final expression of $a_{lm}$ is given by
\bear
a_{lm}(\nu) 
&=&
4 \pi \bar{T}~ \bar{x}_{\rm HI}
(-{\rm i})^l~\int \f{\de^3 k}{(2 \pi)^3} Y^*_{lm}({\bf \hat{k}})
\nline
&\times&
\left[\Delta_{\rm HI}({\bf k}) j_l(k r_{\nu})
- \Delta({\bf k}) j''_l(k r_{\nu}) \right]
\ear

The next step is to calculate the 
the power spectrum
$C_l(\nu_1, \nu_2) \equiv \langle a_{lm}(\nu_1) ~ a^*_{lm}(\nu_2) \rangle$.
The corresponding expression is then given by
\bear
C_l(\nu_1, \nu_2) 
\!\!\!\!\!&=& \!\!\!\!\!
(4 \pi)^2 \bar{T}(z_1)\bar{T}(z_2) ~ \bar{x}_{\rm HI}(z_1)\bar{x}_{\rm HI}(z_2)
\nline
\!\!\!\!\!&\times&\!\!\!\!\!
\int \f{\de^3 k_1}{(2 \pi)^3} \int \f{\de^3 k_2}{(2 \pi)^3}
Y^*_{lm}({\bf \hat{k}_1}) Y_{lm}({\bf \hat{k}_2}) 
\nline
\!\!\!\!\!&\times&\!\!\!\!\!
\left\langle
\left[\Delta_{\rm HI}(z_1, {\bf k_1}) j_l(k_1 r_{\nu_1})
- \Delta(z_1, {\bf k_1}) j''_l(k_1 r_{\nu_1}) \right]
\right.
\nline
\!\!\!\!\!&\times& \!\!\!\!\!
\left.
\left[\Delta^*_{\rm HI}(z_2, {\bf k_2}) j_l(k_2 r_{\nu_2})
- \Delta^*(z_2, {\bf k_2}) j''_l(k_2 r_{\nu_2}) \right]
\right\rangle
\nline
\ear
where we have put back the redshift-dependence into the expressions
for clarity.
Now note that we would mostly be interested in cases where 
$\nu_2 - \nu_1 \equiv \Delta \nu \ll \nu_1$. In such cases, one can
safely assume $\bar{T}(z_2) \approx \bar{T}(z_1)$ and
$\bar{x}_{\rm HI}(z_2) \approx \bar{x}_{\rm HI}(z_1)$.
Furthermore, the terms involving the ensemble averages
of the form
$\langle \Delta \Delta^* \rangle$ can be approximated
as $\langle\Delta(z_1, {\bf k_1}) \Delta^*(z_2, {\bf k_2})\rangle
\approx (2 \pi)^3 \delta_D({\bf k_1 - k_2}) P(z_1, k_1)$
and similarly for terms involving $\Delta_{\rm HI}$.
We can then use the Dirac delta function 
$\delta_D({\bf k_1 - k_2})$ to compute the 
${\bf k_2}$-integral, and thus can write the 
angular power spectrum as
\bear
C_l(\Delta \nu) \!\!\!\!\!&\equiv& \!\!\!\!\! C_l(\nu, \nu + \Delta \nu)
\nline
\!\!\!\!\!&=& \!\!\!\!\!
(4 \pi)^2 ~ \bar{T}^2 ~ \bar{x}^2_{\rm HI}
\int \f{\de^3 k_1}{(2 \pi)^3} 
Y^*_{lm}({\bf \hat{k}_1}) Y_{lm}({\bf \hat{k}_1}) 
\nline
\!\!\!\!\!&\times&\!\!\!\!\!
\left[j_l(k r_{\nu}) j_l(k r_{\nu_2})
P_{\Delta^2_{\rm HI}}(k)
\right.
\nline
\!\!\!\!\!&-&\!\!\!\!\!
\{j_l(k r_{\nu}) j''_l(k r_{\nu_2}) 
+ j_l(k r_{\nu_2}) j''_l(k r_{\nu})\}
P_{\Delta_{\rm HI}}(k)
\nline
\!\!\!\!\!&+&\!\!\!\!\! 
\left. j''_l(k r_{\nu}) j''_l(k r_{\nu_2}) P(k)
\right]
\ear
Using the normalization property of the spherical harmonics
$\int \de {\bf {\hat n}} |Y_{lm}({\bf \hat{n}})|^2 = 1$, one can 
carry out the angular integrals in the above expression, 
and hence obtain the final result (\ref{eq:cl_delta_nu}) 
as quoted in the
main text.

\section{Correspondence between all-sky and flat-sky power spectra}
\label{sec:flat-all}

As discussed in section \ref{sec:flat-sky}, 
we shall mostly be interested in very small
angular scales, which corresponds to $l \gg 1$. 
For high values of $l$, it is most useful to
work in the flat-sky approximation, where 
a small portion of the sky can be
approximated by a plane.
Then the unit vector 
${\bf \hat{n}}$ towards the direction of observation
can be decomposed into ${\bf \hat{n}} = {\bf m} + \boldsymbol{\theta}$,
where ${\bf m}$ is a vector towards the center of the 
field of view and $\boldsymbol{\theta}$ is a two-dimensional
vector in the plane of the sky.

Without loss of generality, let us now
consider a small region around the pole $\theta \to 0$.
In that case the vector $\boldsymbol{\theta}$ can 
be treated as a Cartesian vector with components 
$\{\theta \cos \phi, \theta \sin \phi\}$.
This holds true for any two-dimensional vector on the sky, 
in particular
${\bf U} = \{U \cos \phi_U, U \sin \phi_U\}$.
Then the spherical harmonic components of $T(\nu,{\bf \hat{n}})$ 
[defined in equation (\ref{eq:a_lm})] can be written as
\be
a_{lm}(\nu) \approx \int \de \boldsymbol{\theta} ~ Y^*_{lm}(\theta, \phi) ~ 
T(\nu,{\bf \hat{n}}) 
\label{eq:a_lm_flat}
\e
where we have replaced $\int \de \Omega \rightarrow \int \de \boldsymbol{\theta}$.
Now use the expansion 
\be
{\rm e}^{-2 \pi {\rm i} {\bf U \cdot} \boldsymbol{\theta}}
= \sum_m (-{\rm i})^m J_m(2 \pi U \theta) 
{\rm e}^{{\rm i} m (\phi_U-\phi)}
\label{eq:exp_bessel_2d}
\e
where $J_m(x)$ is the ordinary Bessel function.
Further, we use the approximation for spherical harmonics
\be
Y_{lm}(\theta, \phi) ~ ^{\approx}_{\theta \to 0} ~ J_m(l\theta) 
\sqrt{\f{l}{2 \pi}} ~ {\rm e}^{{\rm i} m \phi}
\e
to write
\be
{\rm e}^{-2 \pi {\rm i} {\bf U \cdot} \boldsymbol{\theta}}
\approx \sqrt{\f{1}{U}} ~ 
\sum_m (-{\rm i})^m Y^*_{2 \pi U, m}(\theta, \phi) 
{\rm e}^{-{\rm i} m \phi_U}
\label{eq:exp_ylm_2d}
\e
Then the two-dimensional Fourier transform of the 
brightness temperature [defined in equation (\ref{eq:T_four})]
will be
\bear
\tilde{T}(\nu, {\bf U}) 
\!\!\!\!\!&=&\!\!\!\!\! 
\int \de \boldsymbol{\theta} ~  
{\rm e}^{-2 \pi {\rm i} {\bf U \cdot} \boldsymbol{\theta}} ~ 
T(\nu,{\bf \hat{n}}) 
\nline
\!\!\!\!\!&\approx&\!\!\!\!\!
\sqrt{\f{1}{U}} ~ 
\sum_m (-{\rm i})^m
{\rm e}^{-{\rm i} m \phi_U}~ 
\int \de \boldsymbol{\theta} ~  Y^*_{2 \pi U,m}(\theta, \phi)~  T(\nu,{\bf \hat{n}}) 
\nline
\!\!\!\!\!&=&\!\!\!\!\!
\sqrt{\f{1}{U}} ~ 
\sum_m (-{\rm i})^m
{\rm e}^{-{\rm i} m \phi_U}~ 
a_{2 \pi U, m}(\nu)
\ear
where we have used the expression (\ref{eq:a_lm_flat}) for $a_{lm}$
in the last part.
This gives a relation between the flat-sky Fourier transform
$\tilde{T}(\nu, {\bf U})$ and and its the full-sky equivalent
$a_{lm}(\nu)$.

Using the above relation, we can
calculate the power spectrum
\bear
\langle \tilde{T}(\nu_1, {\bf U}) \tilde{T}^*(\nu_2, {\bf U'}) \rangle
\!\!\!\!\!&\approx&\!\!\!\!\!
\sqrt{\f{1}{U U'}} ~ 
\sum_{m m'} (-{\rm i})^{m-m'}
{\rm e}^{-{\rm i} m \phi_U}~ {\rm e}^{{\rm i} m' \phi_{U'}}
\nline
\!\!\!\!\!&\times&\!\!\!\!\!
\langle a_{2 \pi U, m}(\nu_1) a^*_{2 \pi U', m'}(\nu_2) \rangle
\ear
Use the definition $\langle a_{lm}(\nu_1) a^*_{l' m'}(\nu_2) \rangle
= C_l \delta_{ll'} \delta_{mm'}$
and the property 
\be
\sum_m{\rm e}^{-{\rm i} m (\phi_U - \phi_{U'})} = 
2 \pi \delta^{(1)}_D(\phi_U - \phi_{U'})
\label{eq:exp_delta_1d}
\e
to obtain
\be
\langle \tilde{T}(\nu_1, {\bf U}) \tilde{T}^*(\nu_2, {\bf U'}) \rangle
= 2 \pi ~ C_{2 \pi U}(\nu_1, \nu_2) ~ \f{\delta_{U U'}}{U} ~ 
\delta^{(1)}_D(\phi_U - \phi_{U'})
\label{eq:T_tilde_sf}
\e
The last step involves writing the right hand side of the above equation
in terms of the two-dimensional Dirac delta function, which
follows from the expansion
\be
\delta^{(2)}_D({\bf U - U'}) =
\int \de \boldsymbol{\theta} ~ 
{\rm e}^{-2 \pi {\rm i} {\bf (U - U') \cdot} \boldsymbol{\theta}}
\e
The exponentials can be written in terms of the 
spherical harmonics using equation (\ref{eq:exp_ylm_2d}):
\bear
\delta^{(2)}_D({\bf U - U'})
\!\!\!\!\!&\approx&\!\!\!\!\!
\int \de \boldsymbol{\theta} ~ 
\sqrt{\f{1}{U U'}} \sum_{m m'} 
(-{\rm i})^{m-m'}
\nline
\!\!\!\!\!&\times& \!\!\!\!\!
Y^*_{2 \pi U, m}(\theta, \phi) Y_{2 \pi U', m'}(\theta, \phi) 
{\rm e}^{-{\rm i} m \phi_U} {\rm e}^{{\rm i} m' \phi_{U'}}
\nline
\ear
Finally use the orthonormality 
property of spherical harmonics 
$\int \de \boldsymbol{\theta} Y^*_{lm}(\theta, \phi) 
Y_{l'm'}(\theta, \phi) = \delta_{ll'} \delta_{mm'}$
and the relation (\ref{eq:exp_delta_1d}) to obtain
\be
\delta^{(2)}_D({\bf U - U'})=
2 \pi ~ \f{\delta_{U U'}}{U} 
\delta_D(\phi_U - \phi_{U'})
\e
Putting the above relation into (\ref{eq:T_tilde_sf}), we obtain
equation (\ref{eq:T_Cl}) used in the final text.

\end{document}